\documentclass[twocolumn,showpacs,showkeys,preprintnumbers,amsmath,amssymb]{revtex4}
\usepackage{graphicx}
\usepackage{dcolumn}
\usepackage{bm}

\begin{document}

\preprint{Regular Article}

\title{High order fractional microwave induced resistance oscillations in 2D systems}

\author{S. Wiedmann$^{1,3}$, G. M. Gusev,$^2$ O. E. Raichev$^{4}$, A. K.
Bakarov,$^2$\footnote{Permanent address: Institute of Semiconductor
Physics, Novosibirsk 630090, Russia}
 and J. C. Portal$^{1,3,5}$}
\affiliation{$^1$LNCMI-CNRS, UPR 3228, BP 166, 38042 Grenoble Cedex 9,
France} \affiliation{$^2$Instituto de F\'{\i}sica da Universidade de
S\~ao Paulo, CP 66318 CEP 05315-970, S\~ao Paulo, SP, Brazil}
\affiliation{$^3$INSA Toulouse, 31077 Toulouse Cedex 4, France}
\affiliation{$^4$Institute of Semiconductor Physics, NAS of Ukraine, 
Prospekt Nauki 45, 03028 Kiev, Ukraine}
\affiliation{$^5$Institut Universitaire de France, 75005 Paris, France}
\date{\today}

\begin{abstract}
We report on the observation of microwave induced resistance oscillations 
associated with the fractional ratio $n/m$ of the microwave irradiation 
frequency to the cyclotron frequency for $m$ up to 8 in a two-dimensional 
electron system with high electron density. The features are quenched at 
high microwave frequencies independent of the fractional order $m$. We 
analyze temperature, power and frequency dependencies of the magnetoresistance 
oscillations and discuss them in connection with existing theories. 
\end{abstract}


\maketitle

\section{Introduction}

In recent years, remarkable effects in two-dimensional (2D) electron
systems have been discovered in the presence of microwave (MW) illumination
and a classically strong transverse magnetic field $B$. It has been observed
that the magnetic-field dependence of the longitudinal resistance in a
high-mobility 2D electron gas exhibits oscillations with a period determined
by the ratio of the radiation frequency $\omega$ to the cyclotron frequency
$\omega_{c}$=$eB/m^{*}$, where $m^{*}$ is the effective mass of electrons \cite{1}.
Later experiments on the samples with ultrahigh mobility have shown that for
a sufficiently high radiation power the resistance minima evolve into "zero
resistance states" (ZRS), where dissipative resistance vanishes \cite{2,3}.
Both the ZRS and the microwave induced resistance oscillations (MIROs) have
attracted much theoretical interest, and several microscopic mechanisms
responsible for these phenomena have been proposed. Two of these mechanisms,
widely discussed in literature, produce similar results concerning the MIRO
periodicity and phase. The first, "displacement" mechanism, \cite{4,5} accounts
for the displacement of electrons along the applied dc field under scattering-assisted
microwave absorption. The second, "inelastic" mechanism, \cite{6} is associated
with a microwave-generated nonequilibrium oscillatory component of the isotropic
part of electron distribution function. The MIROs can also appear owing to the
"photovoltaic" mechanism \cite{7} describing combined action of the microwave
and dc fields on both temporal and angular harmonics of the distribution
function. Finally, it has been predicted that microwave excitation of the
second angular harmonic of the distribution function, which is referred to
as the "quadrupole" mechanism, \cite{7} should lead to an oscillatory 
contribution to the transverse (Hall) resistivity.
Among these mechanisms, the inelastic one is thought to play the dominant role in
experiments at low temperatures $T$ and microwave powers, because its contribution
is parametrically large in comparison to contributions of the other mechanisms.
In particular, the inelastic mechanism overcomes the displacement one by a large
factor $\tau_{in}/\tau_{q} \propto~ T^{-2}$, where $\tau_{in}$ and $\tau_{q}$ are
the inelastic scattering time and quantum lifetime, respectively. It has been
found that the inelastic mechanism satisfactory explains the power and temperature
dependence of resistivity observed in earlier experiments and in more recent
experiments both on single-subband \cite{8} and two-subband 2D systems \cite{9}.

With increasing microwave power, notable MIRO features associated
with fractional ratios
\begin{equation}
\epsilon \equiv \omega/\omega_{c}=n/m,
\end{equation}
where $n$ and $m$ are integers, have been observed \cite{8,10,11,12}. While all
the resonances described by Eq. (1) originate from commensurability of the
cyclotron and microwave frequencies, the fractional MIROs (FMIROs) corresponding
to $m \geq 2$ require multiphoton absorption processes which appear in
the strongly nonlinear regime and can proceed in two ways. The first is a 
simultaneous absorption of several photons (with virtual intermediate states) 
\cite{13,14} and the second is a stepwise absorption of single photons \cite{15}. 
The latter process contributes to resistivity owing to incomplete relaxation of 
the electron system between the absorption events. The MIRO theories accounting 
for these processes \cite{13,14,15} demonstrate progressive appearance of the 
oscillation peaks with increasing fractional denominator $m$ as the microwave 
power increases, which is in agreement with experimental findings \cite{8,10,11,12}. 
The experiment \cite{8} shows the FMIRO features up to $m=4$ ($\epsilon=1/4$).

In this paper we report the observation of oscillatory features which are 
likely attributed to higher-order (up to $\epsilon=1/8$) FMIROs. These oscillations 
are more pronounced at high temperatures ($T \simeq$ 6.5~K), when their observation is not 
hindered by the presence of Shubnikov-de Haas (SdH) oscillations. We analyzed temperature, 
power and frequency dependencies of the magnetoresistance oscillations. For the 
microwave powers used in our experiment, the simultaneous absorption of more 
than two photons is found to be a rare event. Therefore, a more plausible 
explanation for the high-order FMIROs is the stepwise absorption of single 
photons \cite{15}. However, this mechanism also fails to describe observation 
of high-order fractions because of energy conservation restrictions, when 
the gap between the Landau levels exceeds the photon energy. Moreover, we 
have found that the amplitude of FMIROs for {\em all} fractions dramatically 
decreases at high microwave frequencies ($f$~$\geq$~85~GHz). This behaviour 
qualitatively disagrees with the stepwise absorption model \cite{15}, 
which predicts that the threshold frequency strongly depends on the fractional 
value. We believe that this FMIRO quenching effect is very similar
to integer MIRO damping which occurs at higher frequencies \cite{16,17}.

\section{Experiment}

\begin{figure}[h]
\includegraphics[width=9cm]{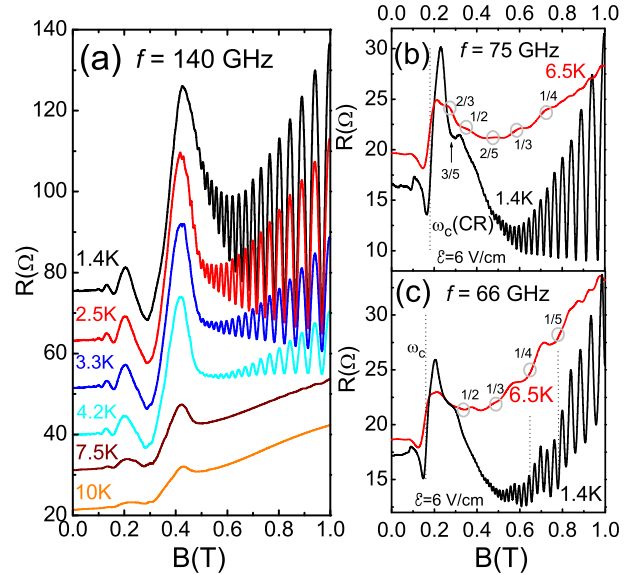}
\caption{\label{fig1} (Color online) (a) Temperature dependent magnetoresistance
$R_{xx}$ of MIROs under microwave irradiation of 140~GHz. The curves 
(except the one for 10 K) are shifted for clarity. (b), (c) 
FMIROs for 75~GHz and 66~GHz at $T$~=~1.4~K and $T$~=~6.5~K. CR (dotted line)
and FMIRO features are marked with circles for corresponding $\epsilon$.}
\end{figure}

The measurements have been performed in GaAs quantum wells (QWs) with well widths of 14~nm
in the presence of microwave irradiation in the range from 32~GHz to 100~GHz
with different intensities of radiation. In contrast to the previous studies
\cite{8,10,11} with low electron density ($< 3.6 \times 10^{11}$ cm$^{-2}$)
and ultrahigh mobility ($\sim 2 \times 10^{7}$ cm$^{2}$/V s), we use high-density
samples from different wafers with moderate mobilities $< 1.8 \times 10^{6}$ cm$^{2}$/V s.
All the data presented below are for a sample with a carrier concentration of $7.8 \times 
10^{11}$~cm$^{-2}$ and a mobility of $1.2 \times 10^{6}$ cm$^{2}$/V s.
The sample was mounted in a VTI cryostat with a waveguide to deliver radiation
down to the sample. The resistance $R=R_{xx}$ was measured by using a standard
low-frequency lock-in technique (13~Hz) under continuous MW illumination.
Whereas previous experiments \cite{8,10,11,12} have been carried out at low
temperature ($T \sim 1$ K), we extended our studies up to 10 K.
Fig. \ref{fig1}(a) presents temperature dependence of MIRO from 1.4~K to 10~K
for a MW irradiation of 140 GHz with 0~dB (highest intensity). For 1.4 K, we
see the cyclotron resonance (CR) peak ($n=1$) as well as integer MIROs with
$n=2,3$. No ZRS are observed in these samples. With increasing temperature,
the MIRO peaks $n=1,2$ survive up to 10~K. For low frequencies ($f <40$ GHz),
the CR peak amplitude starts to decrease rapidly and vanishes for 35~GHz. This
is attributed to the strong suppression of the Landau quantization by disorder
(i.e. to the exponential smallness of the Dingle factors) in the region of $B$ 
corresponding to the CR.

Fig. \ref{fig1}(b) and Fig. \ref{fig1}(c) illustrate magnetoresistance
measurements for 75 and 66~GHz at 1.4~K and 6.5~K, respectivly. 
%<
At 1.4~K we clearly see only one FMIRO feature for each frequency, and 
for 66~GHz we also observe a kind of modulation at the positions where FMIROs are 
precised at a temperature of 6.5~K. 
%>
For 6.5~K, the observations 
at both 75 and 66~GHz reveal clearly pronouced FMIROs. FMIRO
positions $n/m$ are marked according to Eq. (1) with circles for each fractional feature. In order to 
reproduce these experimental results, we performed power and frequency 
dependent measurements up to a MW frequency of 100~GHz.

\begin{figure}[h]
\includegraphics[width=8cm]{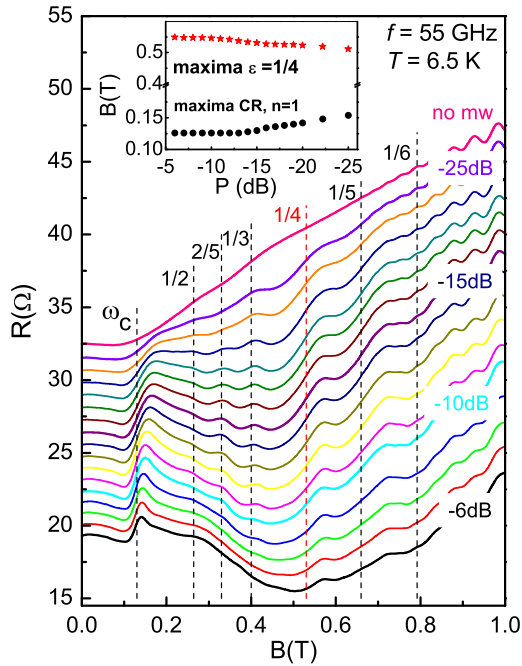}
\caption{\label{fig2} (Color online) (a) Power dependence for 55~GHz
from -6~dB to -25~dB for $f$~=~55~GHz at 6.5~K. The curves (except the 
one for -6 dB) are shifted for clarity. Top trace (no MW) is
the magnetoresistance without microwave irradiation. Positions of the maxima
(CR and FMIROs up to $\epsilon$~=~1/6) are marked by dashed lines. The inset shows
magnetic field dependence of these maxima with decreasing MW power.}
\end{figure}

We now focus on power dependence of FMIROs at 6.5~K, where we find the
best pronounced features, and present our power-dependent measurement
for 55~GHz where we observe more fractional features up to $\epsilon$~=~1/6 
($\epsilon$~=~1/7 is already superimposed by SdH oscillations and not marked in
Fig. \ref{fig2}). In Fig. \ref{fig2} this power dependence is presented for 
MW attenuations from -6~dB to -25~dB. For -6~dB (bottom trace), 
FMIROs $\epsilon$~=~1/4 and $\epsilon$~=~1/5 indicate distinctive local minima and maxima.

With decreasing microwave power, the picture is improved as well as new 
FMIROs in lower magnetic field start to appear ($\epsilon$~=~1/3, $\epsilon$~=~2/5,
$\epsilon$~=~1/2) for -10~dB. The amplitude of these low-field FMIROs increases 
if one further decreases MW power. Looking closer at different FMIROs, it can be 
noticed that down to -25~dB, the amplitude of $\epsilon$~=~1/5 is already suppressed whereas 
$\epsilon$~=~1/4 still exhibits maxima and minima. By plotting the position 
of maxima of CR and FMIRO $\epsilon$~=~1/4 (dotted lines) in magnetic field 
as a function of power, see the inset to Fig. \ref{fig2}, 
we find that for decreasing MW power the FMIRO $\epsilon$~=~1/4 maxima are 
shifted to lower magnetic field in contrast to the evolution of the CR peak
where the maxima shift to a higher magnetic field \cite{6}.
The top trace in Fig. \ref{fig2} is without MW irradiation.
Notice that for 5~K~$<$~$T$~$<$~15~K we see magnetophonon
resonances owing to interaction of electrons with acoustic phonons \cite{18}.

\section{Discussion of the results}

We now discuss the observed frequency, power and temperature dependence of
FMIROs in our samples. The observation of high-order fractional features
can distinguish between the multiphoton model \cite{14} and the
stepwise single-photon mechanism \cite{15}. In fact, absorption of $m > 2$
photons requires a very high microwave power, because the corresponding
contributions into the photoresistance \cite{7,14} should decrease as
$(W_{\pm})^m$, where the dimensionless parameters $W_{\pm}$ ($+$ and 
$-$ stand for two components of microwave radiation with different 
circular polarization) are defined as
\begin{equation}
%2
W_{\pm}= \frac{\tau_q}{\tau_{tr}} \left( \frac{e{\cal E}v_{F}}{\hbar
\omega(\omega_c \pm \omega)} \right)^2,
\end{equation}
and depend on the MW electric field ${\cal E}$, Fermi velocity $v_F$,
and on the ratio of quantum lifetime $\tau_{q}$ to transport time $\tau_{tr}$.
This ratio is usually smaller than 0.1 and is estimated as 1/15 for
our samples. For the MW powers used in our experiments [we estimate
$\cal E$~=~1~$-$~10 V/cm (see Fig. \ref{fig2}) by comparing the suppression of SdH oscillations
with the dc electric field due to the heating effect \cite{9}] and
reported in the previous studies \cite{8,10,11}, the parameters
$(W_{\pm})^m$ are vanishingly small under FMIRO conditions
$\omega_c =m \omega$, except for the case $m=2$ at the highest powers.
The fractional oscillations near $\epsilon$~=1/2 and 3/2 observed
in previous publications in principle can be explained by the
two-photon process \cite{11,12}. However, the features near
$\epsilon$~=1/3, 1/4, 2/3 and 2/5 are hardly ascribed to this
model. From this point of view, FMIRO due to the process of
stepwise absorption of the several photons seems to be more suitable.
Within the framework of this model \cite{15}, the intensity of the
fractional resonances with denominator $m$ is proportional to the
$m$-th power of the squared Dingle factor, $\exp(-2 m \alpha)$, where 
$\alpha=\pi/\omega_c \tau_q$. This explains why the fractional features 
are absent for the low-field side of the cyclotron resonance (the 
disappearance of integer MIRO and a rapid decrease of the CR peak 
amplitude with lowering MW frequency is of the same origin). However, 
it should be also expected that FMIROs damp exponentially faster than 
the integer oscillations due to a decrease of the quantum lifetime with 
increasing temperature \cite{19}. In contrast, we observe that the FMIRO 
features weaken and eventually dissappear at the same temperature as the 
integer MIRO features. It may be argued that this difference is not 
essential after a transition from overlapped to separated Landau levels, 
where $\exp(-\alpha)$ is of the order unity, and many oscillatory 
harmonics contribute into the density of states. Nevertheless, the 
fractional features in the stepwise absorption model are in any case 
more sensitive to broadening or damping factors than the integer features. 

\begin{figure}[h]
\includegraphics[width=9.2cm]{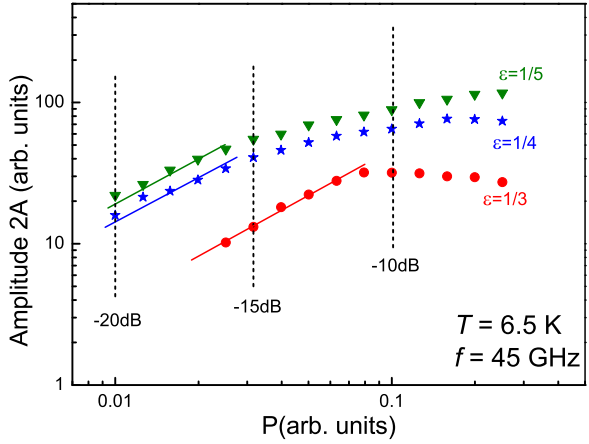}
\caption{\label{fig3} (Color online) Analysis of the power dependence of amplitudes
for CR and FMIROs $\epsilon = 1/3$, 1/4 , and 1/5 for 45~GHz. The amplitude 2A 
(peak to peak) is extracted from the derivative of $R$ as a function of magnetic 
field. The straight lines correspond to a linear power dependence.}
\end{figure}

Let us consider power and frequency dependence as well as the threshold for 
fractional MIROs. It is seen already from Fig. \ref{fig2} that FMIROs are sensitive 
to MW power. Figure \ref{fig3} presents the power dependence of the amplitude 2A, 
extracted from the derivative $d R/d B$, for 45~GHz at 6.5~K for FMIROs 
$\epsilon=1/3, 1/4$, and 1/5. The FMIRO amplitudes show a dependence close to 
linear for weaker MW powers, while for higher powers the dependence becomes 
sublinear and weakly dependent on the power. This behavior is similar to that 
for integer MIRO and is consistent with the stepwise photon absorption 
mechanism \cite{15}. The transition to a weaker power dependence is attributed 
to the saturation effect and confirms the importance of the effect of microwaves 
on the electron distribution function, which is the basic feature of the inelastic 
mechanism of photoresistance \cite{6}. It is worth noting that in the 
region of MW power where we observe FMIROs the power dependence of the 
CR peak amplitude is already sublinear, which means that the saturation regime 
occurs earlier at $\omega \simeq \omega_c$, in agreement with the theory \cite{6}.  
%Likely, the shift
%of the fractional features with increasing power to the direction of lower
%$B$ (Fig. \ref{fig2}), is also related to the nonlinear nature of the fractional MIRO.

\begin{figure}[h]
\includegraphics[width=9cm]{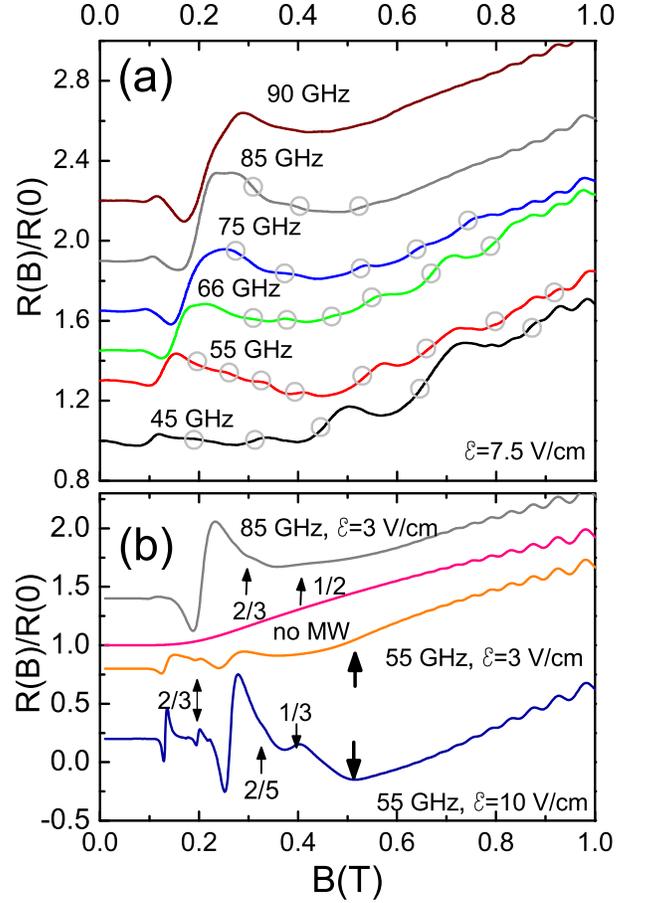}
\caption{\label{fig4} (Color online) Experimental (a) and theoretical (b) magnetoresistance 
for different frequencies at 6.5~K (electric field ${\cal E}$=7.5~V/cm). The curves are 
shifted for clarity. All FMIROs are marked by circles and the corresponding values are listed 
in Table \ref{tab1}. For 90~GHz all the experimental FMIRO features disappear. The theoretical plots 
reveal serveral FMIROs up to $\epsilon$~=~1/3 at ${\cal E}=10$ V/cm but show 
no high-order FMIROs.}
\end{figure}

\begin{figure}[h]
\includegraphics[width=9.2cm]{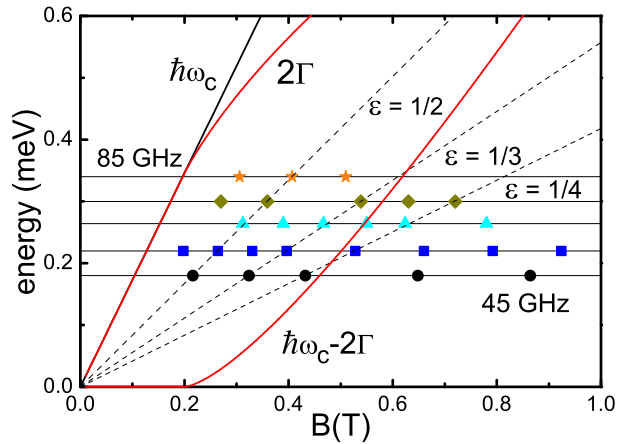}
\caption{\label{fig5} (Color online) Fractions of cyclotron energy, photon energies and 
the threshold energy $\hbar\omega_{c}-2\Gamma$ as functions of the magnetic field.
The positions of the observed fractional features are marked by different 
symbols for each MW frequency. FMIRO features beyond 
the border line can not be described within the mechanism of stepwise 
multiphoton absorption.}
\end{figure}

The measured magnetoresistance as a function of the magnetic field for different 
frequencies up to 90~GHz at 6.5~K is shown in Fig. \ref{fig4}(a). All 
observed FMIROs $\epsilon=n/m$ for the choosen frequencies in Fig. \ref{fig4}(a)
are also presented in Table \ref{tab1} and in Fig. \ref{fig5} with different symbols. 
Apart from FMIROs with $n=1$ and $m=2$, 3,..., which exhibit clear features of 
maxima and minima, we also see weak features at $\epsilon$~=~2/3, 2/5 and 2/7.
No FMIROs for $\omega < \omega_{c}$ ($m < n$) are observed. 
We find experimentally a threshold frequency at $f_{th}$~$\simeq$~90~GHz, 
which is higher than in the previous studies. 

\begin{table}
\caption{\label{tab1}Fractional microwave induced resistance oscillations for all
frequencies in Fig. \ref{fig4}(a) at 6.5~K. The estimated electric field for all
experimental traces is ${\cal E}$=7.5~V/cm.}
\begin{ruledtabular}
\begin{tabular}{lc}
Frequeny (GHz)&FMIROs $\epsilon=n/m$\\
\hline
45 & 1/2, 1/3, 1/4, 1/6, 1/8\footnote{superimposed by SdH oscillations} \\
55 & 2/3, 1/2, 2/5, 1/3, 1/4, 1/5, 1/6, 1/7$^{a}$  \\
66 & 1/2, 2/5, 1/3, 2/7, 1/4, 1/5 \\
75 & 2/3, 1/2, 1/3, 2/7, 1/4 \\
85 & 2/3, 1/2, 2/5 \\
\end{tabular}
\end{ruledtabular}
\end{table}

The results of theoretical calculations accounting for both inelastic and 
displacement mechanisms of photoresistance under both simultaneous 
and stepwise multiphoton processes \cite{13,14,15} are presented in 
Fig. \ref{fig4}(b). The calculation are done using the self-consistent 
Born approximation (SCBA), with $\tau_q=3$ ps and $\tau_{in}/\tau_{tr}=2$, according 
to our estimates for these parameters at $T=6.5$ K. While the theoretical curves 
reproduce general behavior of the magnetoresistance and show low-order 
FMIROs, they do not show high-order FMIROs. The reason for this is the 
following. For observation of the commensurability resonances [see Eq. (1)] 
associated with stepwise single-photon transitions between the adjacent 
Landau levels ($n=1$) in the regime of separated Landau levels ($\alpha<2$ 
according to SCBA) one should satisfy the relation
\begin{equation}
%3
\hbar \omega_{c}-2\Gamma < \hbar \omega < \hbar \omega_{c}+ 2 \Gamma,
\end{equation}
where $\Gamma = \hbar \omega_{c} [ \arccos(1-\alpha)+\sqrt{(2-\alpha)\alpha}]/2\pi$
is the halfwidth of the Landau level in the SCBA. While the upper limit for 
$\omega$ in Eq. (3) is not essential in the range of frequencies and magnetic 
fields we study, the lower limit imposes severe constraints for observation of 
high-order FMIRO, as illustrated in Fig. \ref{fig5}. 

The intersection of the 
photon energy $\hbar \omega$ with the dotted lines marks position of FMIRO at 
$\epsilon=1/m$, while the intersection of the dotted lines with the border 
line $\hbar\omega_{c}-2\Gamma$ shows the threshold energies for corresponding 
fractions. The FMIRO in our theoretical calculations
appear below this threshold, while at the border $\hbar\omega=\hbar\omega_{c}
-2\Gamma$ the magnetoresistance plots have bump-like features clearly visible 
at high MW power and shown by broad arrows in Fig. \ref{fig4}(b).
Note that the quantum lifetime, and, therefore, the halfwidth depends 
on the temperature, thus the high-order FMIROs are favoured at high $T$.
As seen, several features with high order observed at lower $\omega$ are 
not allowed by Eq. (3), especially those for 45 and 55~GHz which occur 
roughly at the same magnetic field. On the other side, the FMIRO with 
$\epsilon=1/2$ is allowed at frequencies above 85 GHz, which also 
disagrees with our observations. It is worth noting that recently it 
has been demonstrated in experiments that the integer MIROs are quenched 
at frequencies above 200 GHz, and practically disappear at higher 
frequencies \cite{16}. Note that in Ref. \cite{17} the mobility is two
times higher than in Ref. \cite{16} which causes the MIRO observation at
240~GHz. The microscopic mechanism of this quenching effect is still unclear. 
We believe that a common theoretical approach to explain the damping effect 
of both fractional and integer MIROs is necessary. 

\section{Conclusion}

In summary, we observe fractional MIROs with ratios $\epsilon=\omega/\omega_{c}=n/m$ 
in a high density two-dimensional electron system for different frequencies 
and relatively high temperatures when SdH oscillations 
are suppressed. Fractional MIROs have a similar temperature dependence to integer 
MIRO, which may indicate that FMIROs are dominated by the inelastic mechanism 
under multiphoton absorption. The transition to sublinear power dependence of FMIRO 
amplitude with increasing microwave power, which we attribute to saturation effect, 
is another argument in favour of the inelastic mechanism \cite{6}. We have analyzed two 
competing models for FMIRO, a simultaneous multiphoton absorption and a stepwise 
absorption of several photons, in application to our data. The multiphoton model 
fails to explain the high-order fractional features because the microwave power 
is not sufficient to generate such multiphoton processes. Concerning the stepwise 
absorption model, both the existence of high-order FMIRO in the absence of 
single-photon transitions between Landau levels and the independence of the threshold 
frequency for FMIRO quenching of the number of fraction disagree with the predictions 
of this model. Therefore we believe that the high-order fractional features require 
another explanation. More experimental work and theoretical effort has to be done in 
order to obtain a complete understanding of this FMIRO phenomenon. 
	
This work was supported by COFECUB-USP (project number U$_{c}$~109/08), CNPq, 
FAPESP and with microwave facilities from ANR MICONANO.

%\bibliography{apssamp}

\begin{references}

\bibitem{1}
M. A. Zudov, R. R. Du, J. A. Simmons, and J. L. Reno, Phys. Rev. B
{\bf 64}, 201311(R) (2001).

\bibitem{2}
R. G. Mani, J. H. Smet, K. von Klitzing, V. Narayanamurti, W. B. Johnson,
and V. Umansky, Nature {\bf 420}, 646 (2002).

\bibitem{3}
M. A. Zudov, R. R. Du, L. N. Pfeiffer, and K. W. West, Phys. Rev. Lett. {\bf 90},
046807 (2003).

\bibitem{4}
V.I. Ryzhii, Fiz. Tverd. Tela (Leningrad) \textbf{11}, 2577 (1969) [Sov.
Phys. Solid State \textbf{11}, 2078 (1970)]	; V. I. Ryzhii, R. A. Surris,
and B. S. Shchamkhalova, Fiz. Tekh. Poluprovodn. \textbf{20}, 2078
(1986) [Sov. Phys. Semicond. \textbf{20}, 1299 (1986)].
\bibitem{5}
A. C. Durst, S. Sachdev, N. Read, and S. M. Girvin, Phys. Rev. Lett {\bf 91}, 086803 (2003).

\bibitem{6}
I. A. Dmitriev, M. G. Vavilov, I. L. Aleiner, A. D. Mirlin, and D.
G. Polyakov, Phys. Rev. B {\bf 71}, 115316 (2005).

\bibitem{7}
I. A. Dmitriev, A. D. Mirlin, and D. G. Polyakov, Phys. Rev. B {\bf
75}, 245320 (2007).

\bibitem{8}
S. I. Dorozhkin, J. H. Smet, K. von Klitzing, L. N. Pfeiffer, and K. W. West
JETP Letters {\bf 86}, 543 (2007).
%[Pis'ma ZhETF, 86, 616 (2007)]
%arXiV:cond-mat/0608633

\bibitem{9}
S. Wiedmann, G. M. Gusev, O.E. Raichev, T. E. Lamas, A. K. Bakarov, and J. C. Portal
Phys. Rev. B {\bf 78}, 121301(R) (2008).

\bibitem{10}
S. I. Dorozhkin, J. H. Smet, V. Umansky, and K. von Klitzing,
Phys. Rev. B {\bf 71}, 201306(R) (2005).

\bibitem{11}
M. A. Zudov, R. R. Du, L. N. Pfeiffer, and K. W. West, Phys. Rev. B {\bf 73}, 041303(R) (2006). 

\bibitem{12}
R.G. Mani, J. H. Smet, K. von Klitzing, V. Narayanamurti, W. B. Johnson, and V. Umansky, Phys.Rev. Lett.
{\bf 92}, 146801 (2004).

\bibitem{13}
X. L. Lei and S. Y. Liu, Appl. Phys. Lett. {\bf 88}, 212109 (2006).

\bibitem{14}
I. A. Dmitriev, A. D. Mirlin, and D. G. Polyakov, Phys.Rev. Lett.
{\bf 99}, 206805 (2007).

\bibitem{15}
I. V. Pechenezhskii, S. I. Dorozhkin and I. A. Dmitriev, JETP Letters {\bf 85}, 86 (2007).

\bibitem{16} S. A. Studenikin, A. S. Sachrajda, J.A. Gupta, and Z. R. Wasilewski,
O. M. Fedorych, M. Byszewski, D. K. Maude, M. Potemski, M. Hilke, K. W. West,
and L. N. Pfeiffer, Phys. Rev. B {\bf 76}, 165321 (2007).

\bibitem{17} R. G. Mani, Appl. Phys. Lett. {\bf 92}, 102107 (2008).

\bibitem{18}
M. A. Zudov, I. V. Ponomarev, A.L. Efros, and R.R. Du, J. A. Simmons and J.L. Reno, Phys.
Rev. Lett {\bf 86}, 3614 (2001).

\bibitem{19} N. C. Mamani, G. M. Gusev, T. E. Lamas, A. K. Bakarov, and O.E. Raichev,
Phys. Rev. B {\bf 77}, 205327 (2008).


\end{references}

\end{document}